\documentclass[graybox]{svmult}

\usepackage{mathptmx}       
\usepackage{helvet}         
\usepackage{courier}        
\usepackage{type1cm}        
\usepackage{makeidx}         
\usepackage{graphicx}        
\usepackage{multicol}        
\usepackage[bottom]{footmisc}

\usepackage[T1]{fontenc}
\usepackage[utf8]{inputenc}
\usepackage[caption=false]{subfig}
\usepackage{booktabs}

\makeindex             

\begin{document}

\title*{Seed selection for information cascade in multilayer networks}
\author{Fredrik Erlandsson, Piotr Bródka, and Anton Borg}
\institute{Fredrik Erlandsson \and Anton Borg \at Blekinge Institute of Technology,\\
Department of Computer Science and Engineering, Sweden \\ \email{fredrik.erlandsson@bth.se}
\and Piotr, Bródka \at Wrocław University of Science and Technology, 
\\Department of Computational Intelligence, Poland}
%
%
\maketitle

\abstract{
Information spreading is an interesting field in the domain of online social media. In this work, we are investigating how well different seed selection strategies affect the spreading processes simulated using independent cascade model on eighteen multilayer social networks. Fifteen networks are built based on the user interaction data extracted from Facebook public pages and tree of them are multilayer networks downloaded from public repository (two of them being Twitter networks). The results indicate that various state of the art seed selection strategies for single-layer networks like K-Shell or VoteRank do not perform so well on multilayer networks and are outperformed by Degree Centrality.
}  

\section{Introduction}
Since the emergence of Network Science \cite{barabasi2016network} one of the most interesting research questions was: How the influence and information spread through the network of social interactions and how to maximize it? \cite{kempe2003maximizing} There are many approaches to maximize the final coverage of the spreading and one of them is selecting proper set of initial seeds which will initialize the process. This set should consist of nodes with the highest combined potential to reach as big portion (in terms of no. of members) of network as possible. Those node are often called Influential users and play an important role in information propagation on online social networks as they have the highest impact on other users in the network.

While the problem of seed selection is quite well investigated in single layered networks with many state of the art methods like K-Shell~\cite{Kitsak2010} or VoteRank~\cite{Zhang2016}. The question is if those approaches will still be the best for multilayer networks which are an relatively new trend in how to model complex networks? \cite{dickison2016multilayer}\cite{salehi2015spreading}
Therefore, in this paper we evaluate four seed selection strategies: Degree Centrality~\cite{barabasi2013network}, K-Shell~\cite{Kitsak2010}, VoteRank~\cite{Zhang2016} and ARL~\cite{erlandsson:2016mdpi} (Section \ref{sub:ss_strategies}), using Independent Cascade Model (ICM)~\cite{Shakarian2015} to simulate the spreading process (Section \ref{sub:icm}) over fourteen multilayer networks are built based on the user interaction data extracted from Facebook public pages (Table \ref{tab:fbnetworks}) and tree multilayer networks downloaded from a public repository (Table \ref{tab:networks}). 

The results are presented in Section \ref{sec:results} and indicate that various state of the art seed selection strategies for single-layer networks like K-Shell or VoteRank do not perform so well on multilayer networks and are outperformed by simple Degree Centrality.

\section{Methods}\label{sec:methods}
This section describes the dataset used in our research and social networks created based on it, the information cascade model and various seed selection methods together with the statistical methods used to evaluate our findings.

\subsection{Dataset and network model}\label{sub:data-network}
The dataset used in this study is a subset of public Facebook pages collected by Erlandsson et.\,al.~\cite{erlandsson2016pokemon} and is publicly available at Harvard Dataverse~\cite{DCBDEP_2017}. The data from these pages were parsed and for each post the corresponding likes and comments were extracted. We considered each page a separate dataset/network. Table~\ref{tab:pages} shows the basic information about investigated 14 Facebook pages.

\begin{table}[!ht]
\centering
\caption{Descriptive information of used pages. The columns nodes and edges show the number of elements for the projected networks created for Comments, and Likes respectively.}\label{tab:pages}
{\scriptsize
\begin{tabular}{lrrrrrrrrr}
\toprule
          Page id &  Posts &   Users &  Comments &   Likes &  C edges$^\ddagger$ &  C nodes$^\dagger$ &  L edges$^\ddagger$ &  L nodes$^\dagger$ &  Interactions \\
\midrule
  1 &     86 &     297 &        50 &     549 &                         31 &                         24 &                   5,200 &                     270 &           685 \\
 2 &    301 &     303 &       227 &     502 &                        130 &                         48 &                     542 &                     157 &         1,030 \\
3 &  1,163 &   2,326 &       499 &   2,161 &                      4,361 &                        273 &                 146,231 &                   1,332 &         3,823 \\
4 &  1,777 &     801 &     1,932 &   4,170 &                      1,770 &                        359 &                   4,996 &                     549 &         7,879 \\
5 &  1,013 &   1,636 &     1,463 &   6,880 &                      3,036 &                        403 &                  85,684 &                   1,502 &         9,356 \\
6 &  5,819 &   5,861 &     1,466 &  25,125 &                      4,832 &                        366 &               2,437,479 &                   5,670 &        32,410 \\
7 &  9,391 &  23,431 &    18,571 &  19,623 &                     11,694 &                      3,462 &                 904,901 &                  14,492 &        47,585 \\
8 &    538 &  13,222 &    11,274 &  36,033 &                    285,095 &                      5,566 &               2,249,954 &                  11,141 &        47,845 \\
9 &  1,607 &  33,004 &    16,398 &  39,914 &                    808,650 &                     10,086 &               5,396,069 &                  26,206 &        57,916 \\
10 &  1,445 &  22,488 &     1,946 &  58,695 &                     11,335 &                      1,219 &              16,109,395 &                  21,626 &        62,086 \\
11 & 14,736 &  37,090 &    26,559 &  44,124 &                    151,619 &                      9,325 &               2,950,437 &                  24,324 &        85,419 \\
12 & 14,159 &  69,424 &    31,209 & 147,710 &                  1,600,003 &                     14,637 &              33,547,079 &                  56,641 &       193,078 \\
13 &  1,187 & 104,558 &    18,568 & 278,173 &                    352,789 &                     11,722 &             100,171,084 &                 100,541 &       297,928 \\
14 & 10,781 &  40,368 &    84,484 & 420,257 &                  2,097,013 &                     14,554 &              49,337,665 &                  36,294 &       515,522 \\
\bottomrule
\multicolumn{9}{l}{$^\dagger$ nodes represent users, disconnected nodes (without edges) have been removed for clarity.} \\
\multicolumn{9}{l}{$^\ddagger$ edges are present if two users have acted on the same post.}
\end{tabular}
}\label{tab:fbnetworks}
\end{table}

\begin{figure}[!ht]
\centering
     \subfloat[]{%
     \includegraphics[width=.5\linewidth]{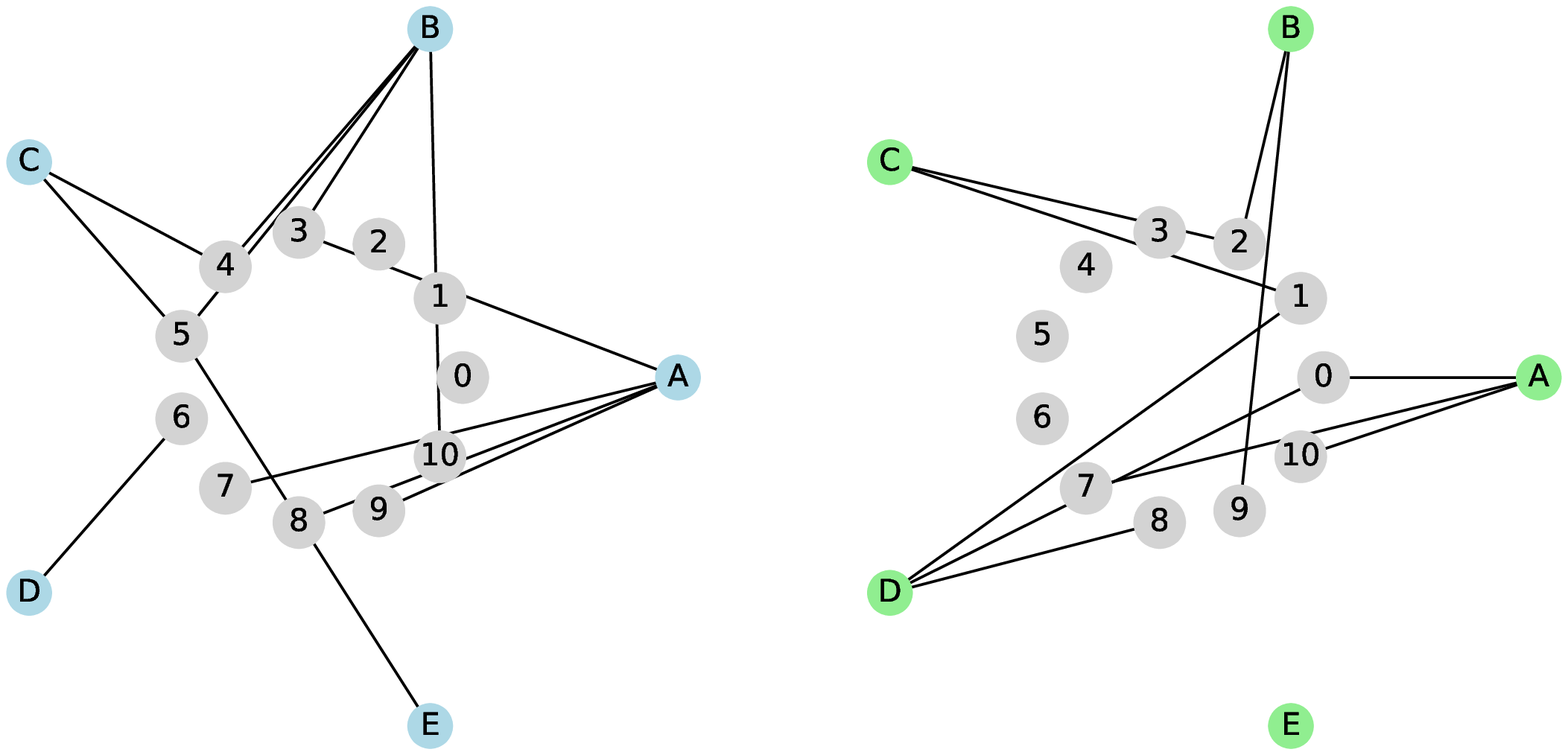}\label{fig:network}}
     \subfloat[]{%
\includegraphics[width=.5\linewidth]{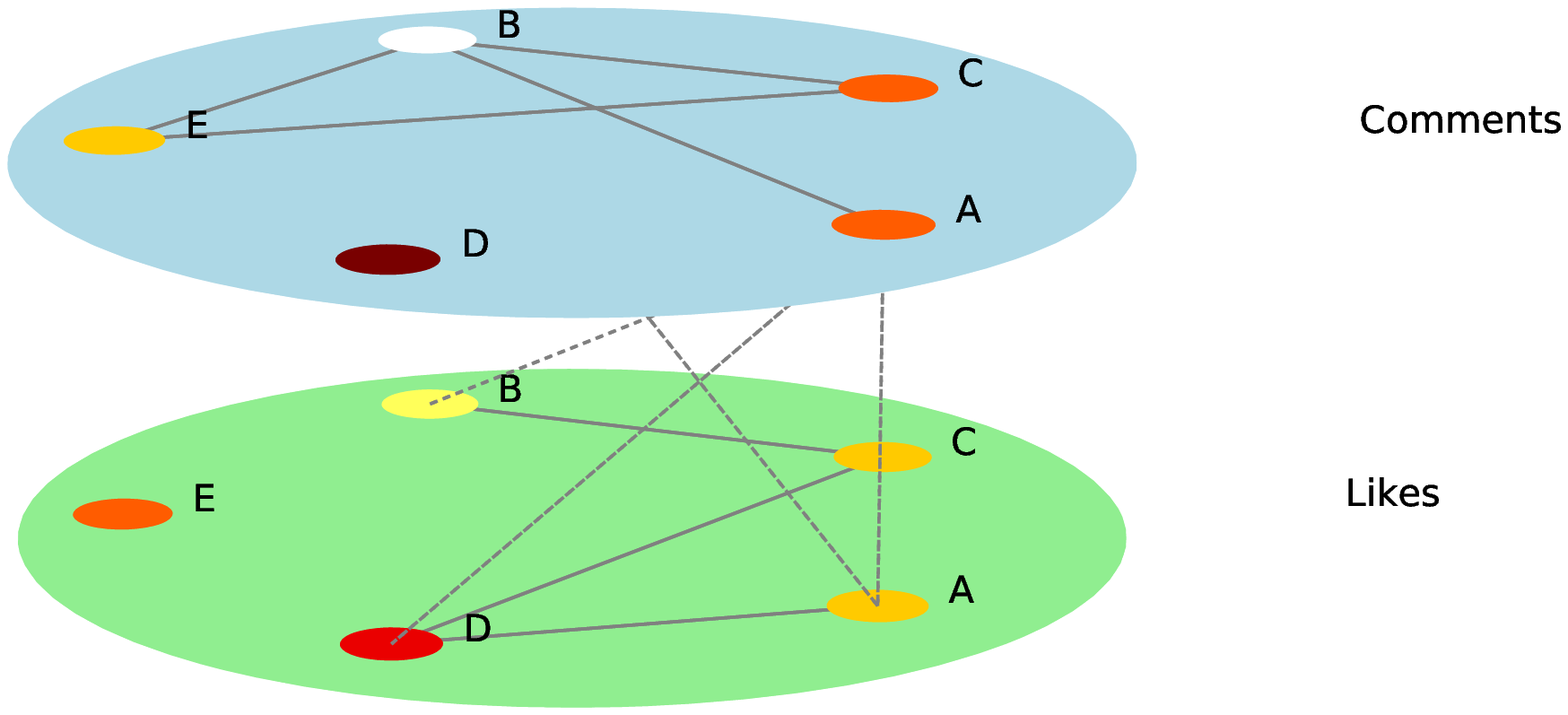}\label{fig:mlnet}}
\caption{A toy example of multiple user interactions as a multilayer network.
\textbf{(a)} The bipartite networks of comments (left) and likes (right). \textbf{(b)} The projected multilayer network from the networks shown in (a).
The green layer represents likes (users like the same post) and the blue layer represents comments (the users are commenting on the same post). Edges between layers (coupling edges) exists between users liking and commenting on the same post. Node $\mathtt{B}$ in the Comments layer represents the initial seed. Nodes and edges with shades ranging from yellow to dark red represents the infection spread.}
\label{fig:stream}\label{fig:networks}
\end{figure}

From each page we build two bipartite networks, one for users' comments and one for users' likes. An example of these two networks are shown in Fig.\ref{fig:network} where the network shown to the left illustrates comments for the users $\mathtt{A-E}$ towards the posts $0-10$, and the network on the right illustrates likes (from the same set of users to the same set of posts). From these two networks we create a multilayer network as shown in Figure~\ref{fig:mlnet}. In the multilayer network the posts have been removed and the interactions towards the post were replaced by direct connection between users interacting with that post. Nodes represents users, and edges between two users indicates that they interacted with the same post, i.e. either they both liked it or they both commented on it. The blue layer represents comments and the green layer represents likes. Each node represent the same user on each layer, i.e., node $\mathtt{A}$ in the Comments layer is the same user as node $\mathtt{A}$ in the Likes layer.

To complement that and to ensure that our findings are not a result of some Facebook properties or the way in which we have prepared our networks we have added to our experiments three social networks from an open repository\footnote{http://deim.urv.cat/~manlio.dedomenico/data.php}, shown in Table~\ref{tab:networks}. Please note that in order to be able to compare the results if some network has more than two layers we are using just two of them. 

\begin{table}[!ht]
\centering
\caption{Descriptive information of used networks.}\label{tab:networks}
{\scriptsize
\begin{tabular}{llrrrrrrrr}
\toprule
                          Id &                         Name &   Users &  Interactions &  L1 nodes &  L1 edges &  L2 nodes &  L2 edges & Source\\
\midrule
 15 &     Pedgett Florentine Families &      15 &            35 &        15 &        20 &        11 &        15 & \cite{action1993rise}\\
16 &     Moscow Athletics 2013 &  88,804 &       197,329 &    74,688 &   104,148 &    46,821 &    89,498 &\cite{omodei2015characterizing}\\
17 &     Marthin Luther King 2013 & 327,707 &       378,462 &   288,738 &   291,083 &    79,070 &    82,987 & \cite{omodei2015characterizing}\\
\bottomrule
\end{tabular}
}
\end{table}

\subsection{Independent Cascade Model}\label{sub:icm}
In this study the Independent Cascade Model (ICM)~\cite{Shakarian2015} was used for modeling information spreading. ICM requires a set of activated nodes at the beginning (seeds) and runs over a limited number of diffusion steps where recently activated nodes has one chance to activate each of its neighbors with the currently configured activation probability.
Thus, if node $\mathtt{A}$ is activated on step 3 it can activate it's neighbors only in step 4, but not in the following steps. 
In our case we ran experiments with the activation probability 1\%. This value was selected due to the high average node degree (edges connected to each node) in the multilayer network.
We have limited the ICM to 10 diffusion steps as the number of activated nodes converge here, and the number of seeds to be 1\% of the number of nodes in the network. 
Here 1\% means that we identify 1\% seeds from each layer and then iteratively select one seed from each layer until we have 1\% in total. Say that we have $\mathtt{\{A,B,C,D\}}$ from layer 1 and $\mathtt{\{E,F,G,H\}}$ from layer 2. This will result in the following seeds $\mathtt{\{A,E,B,F\}}$.
The used implementation of ICM activates nodes separately per layer on each step and the set of activated nodes is updated after each step by summing activated nodes sets from each layer. I.e., diffusion is computed fist for the \emph{Likes} layer and then for the \emph{Comments} layer, if node $\mathtt{B}$ is activated in the \emph{Likes} layer in the current step the node $\mathtt{B}$ will also be activated on the \emph{Comments} layer before continuing to the next diffusion step.

Figure~\ref{fig:mlnet} illustrates the spreading process for our toy example. The initial seed is the node $\mathtt{B}$ in the Comments layer (shown in white). As we are considering each node to be infected after each infection step we let node $\mathtt{B}$ in the Likes layer to also be infected (illustrated as yellow). After the first infection step the node $\mathtt{E}$ in the Comments layer and nodes $\mathtt{C}$ and $\mathtt{A}$  in the Likes layer activated. Resulting in the nodes $\mathtt{C}$ and $\mathtt{A}$ in the Comments layer and node $\mathtt{E}$ in Likes layer also to become infected. In step two the node $\mathtt{D}$ in the Likes layer is activated, thus activates node $\mathtt{D}$ in the Comments layer. With this model and toy network all nodes are activated after just two steps.

\subsection{Seed selection}\label{sub:ss_strategies}
Influential users or, activation seeds were selected using three network based state of the art methods \emph{Degree Centrality}~\cite{barabasi2013network}, \emph{K-Shell}~\cite{Kitsak2010}, and \emph{VoteRank}~\cite{Zhang2016}, together with a machine learning method, \emph{ARL}~\cite{erlandsson:2016mdpi} as an efficient and accurate method for ranking users on social networks. We also included a \emph{Random} sample of seeds as a baseline. All of the investigated methods for seed selection are ranking methods and we select the top nodes with highest rank to use as seeds for the ICM.

Degree Centrality is a network measure which indicates how many connections with the rest of network each node has, it has the advantage of being easy to compute once the network is created. 
K-shell is a measure that is determined using shell decomposition. The highest K-Shell number is considered to be the ``core'' of the network. To efficiently rank seeds we combined the K-Shell rank with degree. By doing this we have created hybrid measure which eliminate K-Shell disadvantage i.e. it does not have enough granularity and multiple nodes can belong to the same K-Shell thus one would have to choose the seeds randomly. 
VoteRank selects seeds iteratively by letting each nodes' neighbors vote using a penalized model where nodes close to an already selected seed will have a decreased voting score/power, and already selected node will not have voting rights.

Using ARL to identify seeds have the advantage of not requiring creating the network before identifying influential users. The chosen ARL algorithm Eclat~\cite{10.1109/69.846291} also have the advantage of being able to reduce the dataset by using a threshold, saying that a user must be active on at least a predefined number of posts to be included in the computation. 
In this work we do not use a fixed value of this threshold, instead we start with an relatively large number and decrease this number until we hit a computational limit and then we use the lowest successful limit and return the computed result. The reported timings shown in Fig.\ref{fig:timings} illustrate the computation time for the final selected threshold as in a real world setting this threshold value will be configured before running the ARL algorithm. The major time consumption of ARL is in building list of when users appear together. For example, user $\mathtt{A}$ is active on posts 1, 2 \& 3, user $\mathtt{B}$ is active on posts 2 \& 3 et cetera. A typical rule is $\mathtt{\{A, B\} \Rightarrow C}$, i.e. if $\mathtt{A}$ and $\mathtt{B}$ appears so will $\mathtt{C}$. A limitation of ARL is that it only ranks a subset of the users. The ranked users are then used as seeds for ICM. The ranked number of seeds identified by ARL is used by VoteRank to limit the number of seeds computed and to compare the models fairly.

\subsection{Statistical evaluation}\label{sub:eval}
The coverage is used as an evaluation metric at each step in the ICM. As such, it is possible at each step to measure how quickly the information spreads. Most research today uses the final coverage as the primary evaluation metric~\cite{Kitsak2010,Zhang2016,zhao2013identifying}. However, as can be seen in Figure~\ref{fig:activation}, the coverage tend to stabilize over seed selection methods after a certain amount of steps. As such, there are drawbacks to using the final coverage as evaluation metric. First, there is always a chance that different algorithms converge to the same final coverage. Second, evaluating the mean or the median of the coverage will also give misleading measurements, as it doesn't take into account the development rate of the coverage. Consequently, evaluating only on the final coverage is inadvisable. 

As such, in this study the primary evaluation metric is the area under curve of coverage (AUC), i.e. how much area will there be under the coverage curve. A larger area denotes a faster rise in coverage, a higher coverage, or both. In this study is the AUC normalized based on the number of diffusion steps computed.

The AUC captures the development of the coverage over the steps in the ICM. Consequently, comparing the AUC allows the comparison of the methods performance on pages. The AUC is calculated using the \emph{MESS} R-package.  It should be noted that the AUC is not to be confused with the AUROC (Area Under Receiver Operating Characteristics curve), which is often colloquially referred to as AUC.

To investigate whether any statistical significant difference exist between the different methods, the Friedman test is used~\cite{Demsar:2006un}. The Friedman test is a non-parametric test that evaluates different treatments (in this case different seed selection algorithms) over multiple datasets. A non-parametric test is chosen over a parametric as normality cannot be assumed over the different datasets. As the test only detects whether a statistical significant difference exists, and not where the difference exists, a post-hoc test is necessary to determine where the difference is located. The Nemenyi test is used as a post-hoc test~\cite{Demsar:2006un}.

\section{Results}\label{sec:results}
We have run experiments for eighteen multilayer networks, 1\% activation probability, 1\% of nodes as initial seeds and five seed selection strategies. This resulted in 90 combinations of experiment parameters. For each combination we run 10 simulations of spreading process using Independent Cascade Model (ICM). The results show that selecting seeds with high Degree Centrality performs the highest activation coverage and also is the simplest and thus fastest method for seed selection.

\begin{figure}[!ht]
\centering
\subfloat[Information cascade on the multilayer network (Likes and Comments)  with activation probability 1\%. \label{fig:activation:ml1}]{%
\includegraphics[width=.45\linewidth]{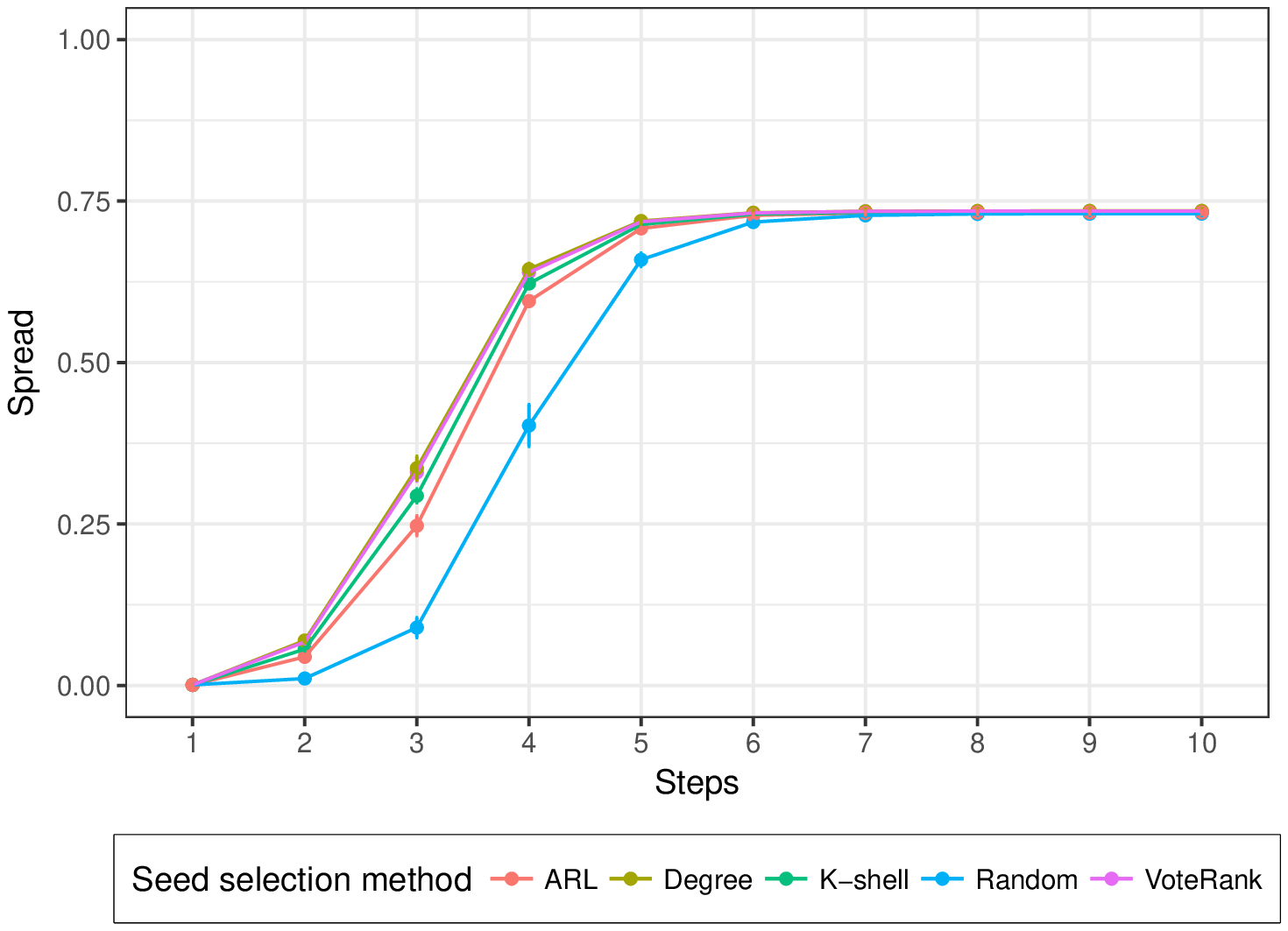}
}
\quad
\subfloat[Information cascade on the multilayer network (Likes and Comments)  with activation probability 10\%.\label{fig:activation:ml10}]{%
\includegraphics[width=.45\linewidth]{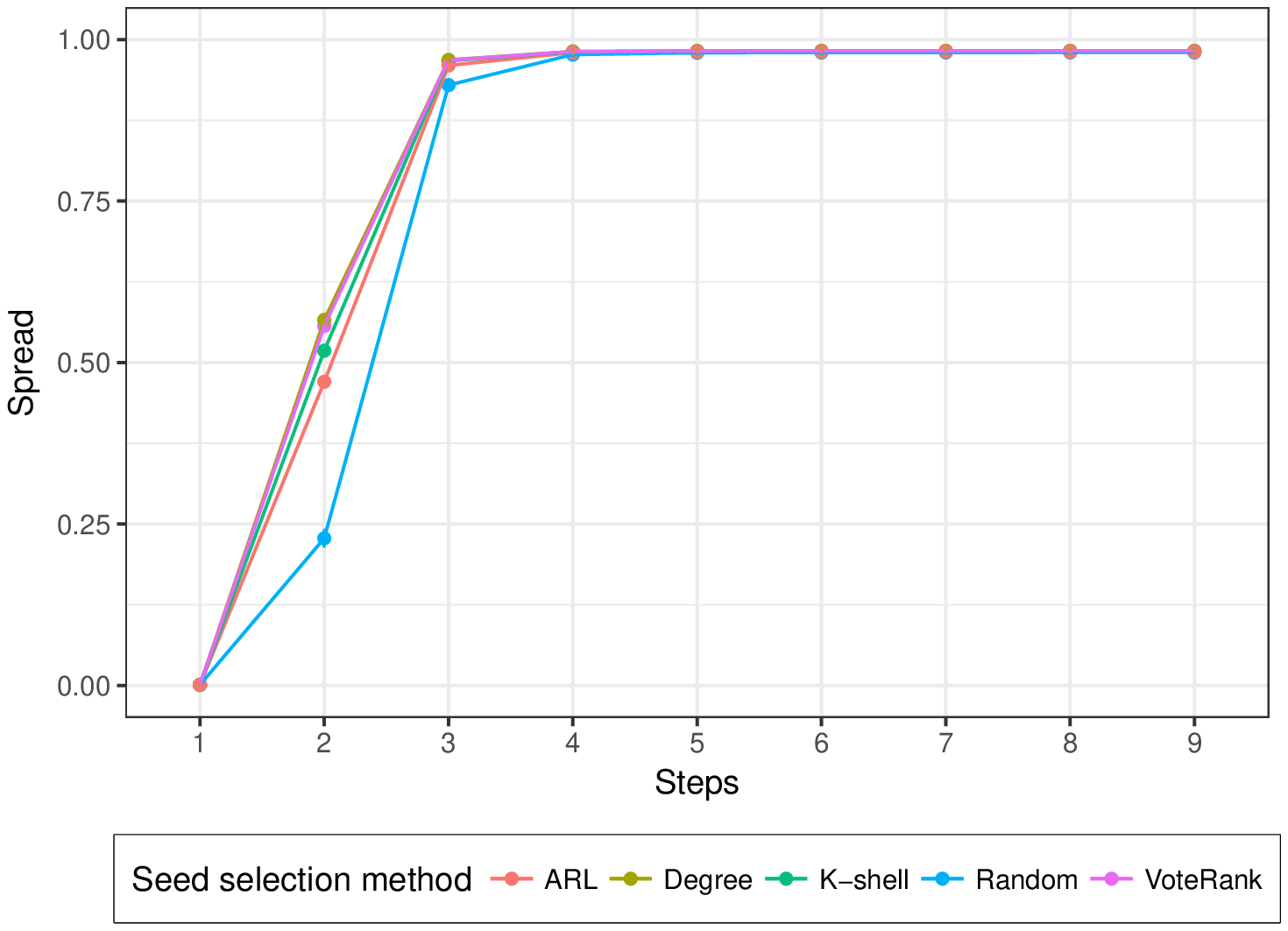}
}

\subfloat[Information cascade on the single layer network (Comments)  with activation probability 1\%. \label{fig:activation:sl1}]{%
\includegraphics[width=.45\linewidth]{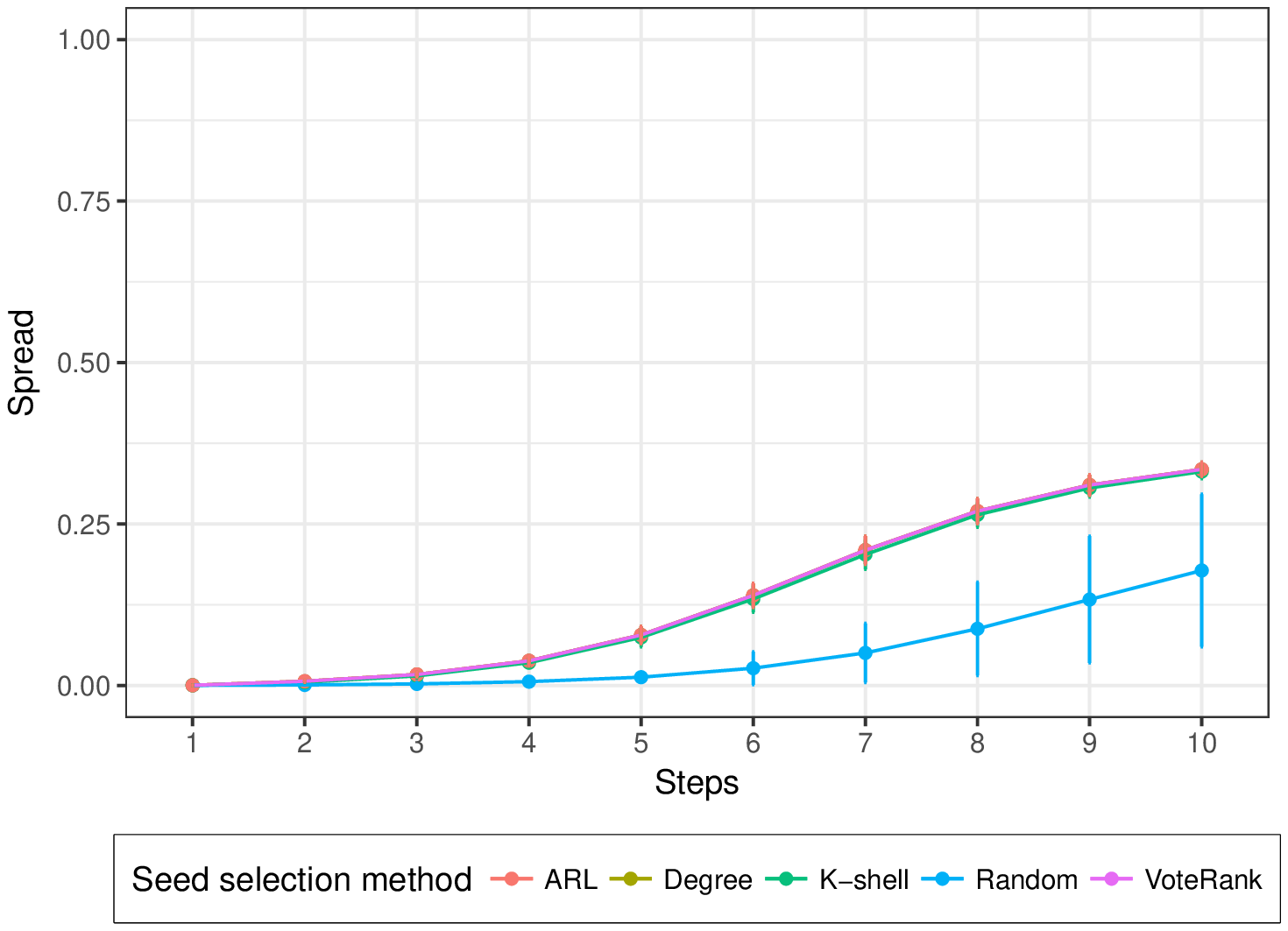}
}
\quad
\subfloat[Information cascade on the single layer network (Comments)  with activation probability 10\%.\label{fig:activation:sl10}]{%
\includegraphics[width=.45\linewidth]{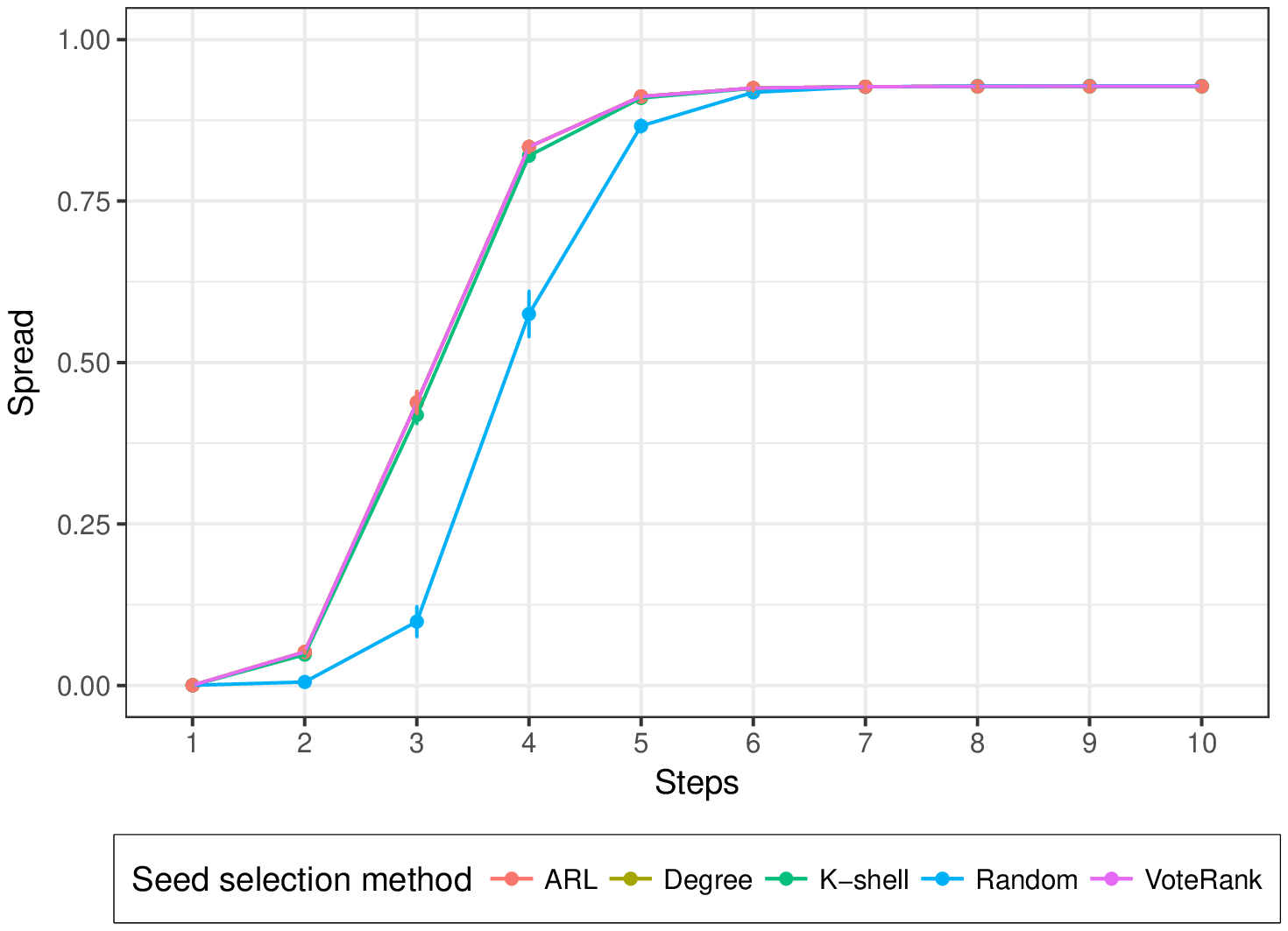}
}

\caption{Activation results for different Activation Probabilities on Multilayer and Single Layer networks for the interactions on the page 7.}\label{fig:activation}
\end{figure}

To illustrate how different activation probabilities and how ICM behaves in both single- and multilayer networks we ran ICM on one of the pages with different settings. Figure~\ref{fig:activation} shows the spreading process for the page no. 8 for different activation probabilities (1\% for Fig~\ref{fig:activation:ml1} and Fig~\ref{fig:activation:sl1}, and 10\% for Fig~\ref{fig:activation:ml10} and Fig~\ref{fig:activation:sl10}) and two different network types. This two types are a multilayer network created from the users' Comments (first layer) and Likes (second layer), shown in Fig~\ref{fig:activation:ml1} and Fig~\ref{fig:activation:ml10}; and a single layer network created from the users' Comments, shown in Fig~\ref{fig:activation:sl1} and Fig~\ref{fig:activation:sl10}. Please note that the plots for the multilayer graph reaches higher coverage faster than the plots for the single-layer graph as the multilayer graph is more dense, see {Table}~\ref{tab:pages} for more information.

\subsection{The final coverage for various seed selection methods}
\begin{figure}[!htbp]
\centering
\includegraphics[width=.95\linewidth]{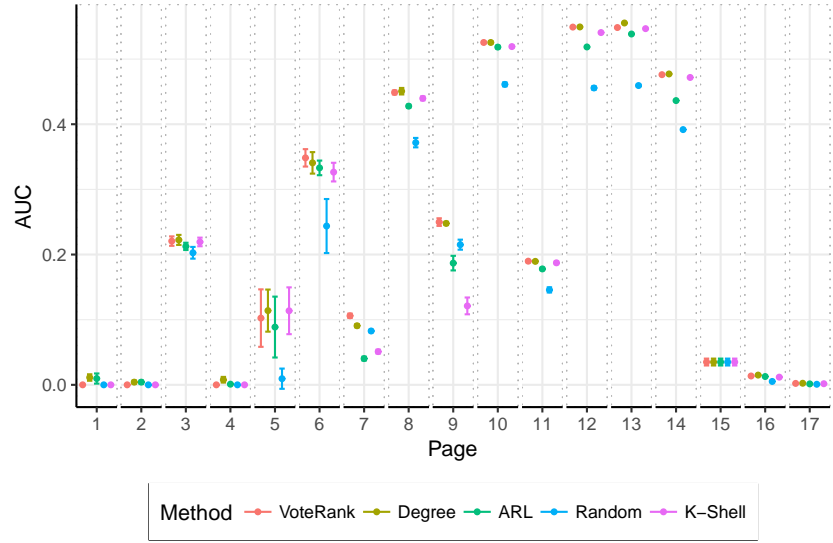}
\caption{Mean AUC of the activation coverage for different seed selection methods. The AUC is calculated on 10 steps and 1\% activation probability of the activation coverage from ICM in a multilayer network. }\label{fig:auc_methods}
\end{figure}

Figure~\ref{fig:auc_methods} shows the resulting mean AUC for the 17 pages investigated. The relatively low AUC is due to some of the multilayer networks have many connected components and that the seeds are just selected from a few of these components.

The Friedman found significant differences between the seed selection methods over the pages ($\chi^2= 43.333$, $df=4$, $p=8.824e^{-09}$), with respect to activation coverage with an activation probability of 1\%. The Nemenyi post-hoc test, presented in Table~\ref{tab:statTest:algo}, shows statistical significant differences between Degree Centrality and a Random sample, ARL, and K-Shell. Further, There were also a statistical significant difference between VoteRank and a Random sample when comparing the AUC.

\begin{table}[!ht]
\centering
\caption{Nemenyi post-hoc test for detecting statistically significant differences between the different seed selection methods with an $AP=1\%$ with respect to the mean AUC of activation coverage.}\label{tab:statTest:algo}
\begin{tabular}{rccccc}
\toprule
\multicolumn{1}{l}{}&\multicolumn{1}{c}{Degree}&\multicolumn{1}{c}{Random}&\multicolumn{1}{c}{ARL}&\multicolumn{1}{c}{K-Shell}&\multicolumn{1}{c}{VoteRank}\tabularnewline
\midrule

Degree&---&3.059&2.088&2&0.794\tabularnewline
Random&$^*$, $^{**}$&---&0.971&1.059&2.265\tabularnewline
ARL&$^*$&&---&0.088&1.294\tabularnewline
K-Shell&$^*$&&&---&1.206\tabularnewline
VoteRank&&$^*$, $^{**}$&&&---\tabularnewline

\bottomrule
\end{tabular}
\\
\footnotesize $^{*}$: Significant difference at $p<0.01$ Critical Difference: $1.765$\\
\footnotesize $^{**}$: Significant difference at $p<0.001$ Critical Difference: $2.103$
\end{table}

As such, the results indicates that Degree Centrality perform significantly better than the other seed selection methods (except VoteRank), i.e. the AUC for this method were significantly larger than for the other seed selection methods in general. Further, VoteRank is significantly better than a Random sample. Interestingly there is no significant difference between a Random sample and either ARL or K-Shell, i.e., selecting seeds using these methods were not statistically better than selecting seeds at random.

\subsection{Time complexity of seed selection methods}
\begin{figure}[!htbp]
\centering
\includegraphics[width=.95\linewidth]{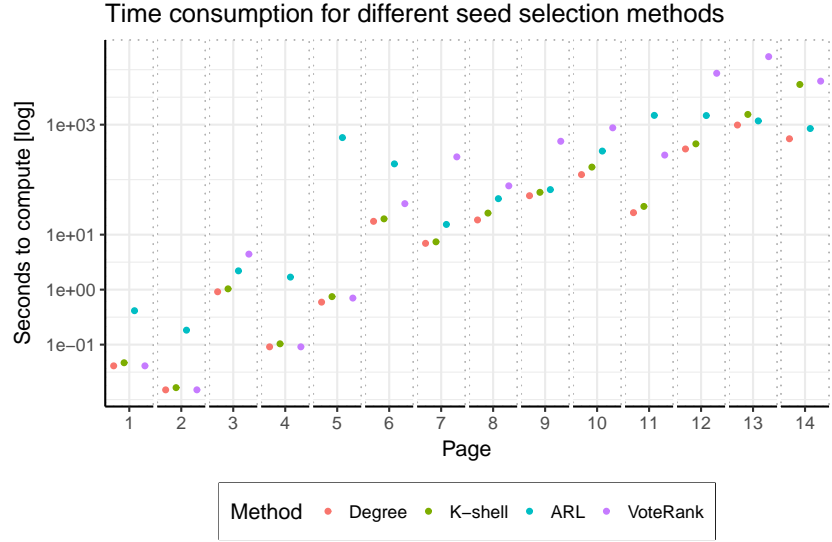}
\caption{Time consumption for different seed selection methods. Included is the preprocessing step of building the graphs. It should be noted that the Y-axis is logarithmic and the top value is around 30.000 seconds.}\label{fig:timings}
\label{fig:time}
\end{figure}

Figure~\ref{fig:timings} show the time complexity of the investigated pages for seed selection with the four different methods. Both VoteRank and ARL are slower than the other methods. The execution time for each page in Fig.~\ref{fig:timings} is an average from ten runs, and the error bars are indicating the standard deviation.

For the three network based seed selection algorithms (Degree, K-Shell and VoteRank) the major time complexity consumer shown in Fig.\ref{fig:timings} is the network creation from our dataset. 
On the other hand, the major time consumer for the ARL method is the building of item sets. Further more, we only calculate the same number of seeds for VoteRank as the ARL method identified, while for Degree and K-Shell we calculated the ranking for the whole network.

A Friedman test shows significant differences between seed selection methods ($\chi^2=29.314$, $df = 3$, $p=1.923e^{-06}$). The Nemenyi post-hoc test found significant differences between Degree Centrality  and all other methods when comparing time complexity for seed selection, e.g. Degree performed significantly faster than the other methods. K-Shell, VoteRank, and ARL are significantly slower than Degree Centrality and there is no internal significant difference between these methods, see Table~\ref{tab:statTest:timings}.

\begin{table}[!htpb]
\centering
\caption{Nemenyi post-hoc test for detecting statistically significant differences between the different seed selection methods with respect to time complexity.}\label{tab:statTest:timings}
\begin{tabular}{r c c c c}
\toprule
& Degree & K-Shell & ARL & VoteRank \\
\midrule
Degree&---&1.429&2.286&2.286\tabularnewline
K-Shell&$^*$&---&0.857&0.857\tabularnewline
ARL& $^*$, $^{**}$&&---&0\tabularnewline
VoteRank& $^*$, $^{**}$&&&---\tabularnewline
\bottomrule
\end{tabular}
\\
\footnotesize $^*$: Significant difference at $p<0.05$ Critical Difference: $1.254$\\
\footnotesize $^{**}$: Significant difference at $p<0.001$ Critical Difference: $1.832$\\
\end{table}

\section{Conclusion}
We have evaluated five seed selection strategies to see how they affects information cascade in multiplex networks. The evaluation was made on 14 public pages on Facebook, two datasets with Twitter data, and one dataset describing Florentine families in the Renaissance. 

The results show that Degree Centrality and VoteRank performs best for seed selection in multiplex networks. The results show that although ARL can be used for seed selection in an information cascade setting it is not preferred as it performs equally as a Random sample. Further, Degree Centrality is significantly faster than the other methods.
If we take into consideration both time complexity and final number of activated users the Degree Centrality is the most optimal seed selection strategy for all tested networks. 

\section*{Acknowledgement}
This work was partially supported by The Polish National Science Centre, the decision no. DEC-2016/21/D/ST6/02408; the European Union’s Horizon 2020 research and innovation programme under the Marie Skłodowska-Curie grant agreement No. 691152 (RENOIR) and the Polish Ministry of Science and Higher Education fund for supporting internationally co-financed projects in 2016-2019 (agreement no. 3628/H2020/2016/2).

\bibliographystyle{spmpsci}

\bibliography{cascade}

\end{document}